\documentclass[%
 reprint,
groupedaddress,
 amsmath,amssymb,
 aps,
]{revtex4-2}

\usepackage{graphicx}
\usepackage{slashed}
\usepackage{dcolumn}
\usepackage{bm}

\begin{document}

\preprint{APS/123-QED}

\title{Optical Effects in an Electric Dipole Spin Polarised Relativistic Quantum Plasma} 

\author{V. M. Demcsak}
 \email{victor.demcsak@sydney.edu.au}

\author{D. B. Melrose}%
 \affiliation{%
School of Physics, The University of Sydney, NSW 2006, Australia
}%

\date{\today}

\begin{abstract}
\noindent The covariant, spin-dependent response tensor for an electric dipole moment polarized electron gas (statistical distribution of electrons and positrons) is calculated using the formalism of quantum plasmadynamics. A simultaneous eigenfunction of both the Dirac Hamiltonian and the electric moment spin operator is constructed. Expressions for the electric moment states and the corresponding vertex functions are derived. It is shown that when the distribution of momenta is isotropic, the spin dependent response of an electric moment dependent quantum plasma is identically zero. The response is non-zero in the presence of a streaming motion perpendicular to the axis if the electron and positron distributions are different. The response has the same form as for a plasma with a nonzero, cross field current when $\langle p_{i} \rangle \neq 0$. This quantum relativistic correction is used to identify the dispersion equation for an electric moment spin polarised plasma with a streaming cold plasma background. In particular, the natural modes of an electric moment dependent quantum plasma exhibit elliptical polarisation. This is in contrast to a magnetic moment dependent electron gas, which is gyrotropic, or a helicity dependent electron gas which is optically active. The different responses, due to quantum plasmas being spin polarised by different relativistically acceptable spin operators, does not appear in other approaches to quantum plasma theory. 
\end{abstract}

\maketitle

\section{Introduction}
The linear response 4-tensor, $\Pi^{\mu\nu}(k)$, relates the 4-current, $J^\mu(k)$, induced by a test field with 4-potential $A^\nu(k)$ in the plasma, where the argument $k$ denotes the components of the wave 4-vector $k^\mu=[\omega,\vec{k}]$, and where we use units with $c=\hbar=1$. The charge-continuity and gauge-independence relations imply $k_\mu\Pi^{\mu\nu}(k)=0$, $k_\nu\Pi^{\mu\nu}(k)=0$. We use the formalism of quantum plasmadynamics (QPD) [1-3] to calculate $\Pi^{\mu\nu}(k)$ for a spin-dependent electron-positron plasma. The generalization of quantum electrodynamics (QED) to QPD involves replacing the electron propagator in vacuo by the electron propagator statistically averaged over the particle occupation number, $n^\epsilon_s(\vec{p})$, with $\epsilon=+1$ for electrons and $\epsilon=-1$ for positrons, and where $s=\pm1$ denotes the eigenvalues of the spin operator. The choice of spin operator is restricted by the requirement that it commute with the Dirac Hamiltonian, so that $s$ is a constant of the motion, and by the requirement that it be a 4-tensor, so that the theory is covariant. The
 form of $\Pi^{\mu\nu}(k)$ depends on the choice of spin operator and on the dependence of $n^\epsilon_s(\vec{p})$ on $s$. Acceptable choices include the helicity and magnetic-moment operators [11]. Here we assume that the spin operator is the electric-moment operator, and we compare the results with those for the other two choices.
\\
\\
Apart from QPD, which is an operator-based theory, three other approaches have been used to describe the response of a relativistic quantum plasma. Quantum fluid theory (QfT) \cite{haas2011quantum} is based on a fluid description of the plasma; plasma kinetic theory (PKT) \cite{bonitz2016quantum} and Dirac-Heisenberg-Wigner theory (DHW) \cite{bialynicki1991phase,vasak1987quantum, hebenstreit2010schwinger} are based on the Wigner function \cite{wigner1997quantum}.

 DHW theory has been used to investigate the Dirac vacuum, in the context of quantum kinetic theory, in a non-manifestly covariant way \citep{bialynicki1991phase}. The DHW formalism \cite{bialynicki1991phase,vasak1987quantum, hebenstreit2010schwinger} requires the Hartree approximation, this is not relevant in QPD. Existing Wigner-function-based theories do not differentiate between spin operators, implicitly ignoring the requirement that a spin operator commute with the Hamiltonian \cite{melrose2020quantum, demcsak2021response}. It is not clear how one may include different choices of spin operators in Wigner function theories of relativistic quantum plasmas. Demcsak and Melrose \citep{demcsak2021response} demonstrated that differences between acceptable spin operators have observable effects in relativistic quantum plasmas. Specifically, when an electron gas (statistical distribution of electrons and positrons) is spin polarised by the magnetic moment operator the plasma exhibits gyrotropy. If the helicity operator is used instead, the response corresponds to an optically active plasma \cite{melrose2003optical}.
\\
\\
The Schwinger effect (electron-positron pair production), in the context of PKT and DHW theory, where an electric field is present, has been studied in \cite{blinne2014pair, blinne2016comparison, hebenstreit2010schwinger}.
\\
\\
In this article we identify the response tensor for an arbitrary unmagnetised relativistic quantum plasma (statistical distribution of electrons and positrons) that is spin polarised by the electric moment operator, assumed to be directed along the $z$-axis. A simultaneous eigenfunction of the Dirac Hamiltonian and the electric  moment operator is identified. This is used to evaluate the corresponding vertex function, which is then used to calculate the spin dependent part of the response tensor. The response is nonzero if there is a streaming motion perpendicular to the axis, such that $\langle p_x\rangle\ne0$ or $\langle p_y\rangle\ne0$, but only if the electron and positron distributions are different. It is shown that this response tensor takes the same form as that of a plasma having a nonzero cross field current when $\langle p_{y} \rangle \neq 0$. An electric moment spin polarised quantum plasma having an isotropic distribution (of momenta) is shown to have no spin dependent contribution to the linear response 4-tensor. 
\\
\\
The difference between the responses of a helicity dependent, magnetic moment dependent and an electric moment dependent relativistic quantum plasma are discussed later in this article.

\section{Response of an unmagnetized electron gas} \label{213367}
In QPD the 4-potential $A^{\mu}(k)$ describes an arbitrary the disturbance in the medium, and the induced 4-current $J^{\mu}(k)$ describes the response, and the linear response is the linear term in an expansion of this current in powers of the disturbance. Here the argument $k$ is the wave 4-vector $k^{\mu}=(\omega, \vec{k})$. The linear response tensor $\Pi^{\mu\nu}(k)$ relates the linear response and the disturbance.  
The linear response tensor gives a complete description of the electromagnetic properties of a given medium. The physical requirements of a medium are given by the mathematical constraints on the response tensor. In particular the 4-response tensor must satisfy charge continuity and gauge invariance. These conditions are identified respectively as: 
\begin{equation}
k_{\mu} \Pi^{\mu \nu}(k)=0, \;\; k_{\nu}\Pi^{\mu \nu}(k)=0.
\label{gaugecharge}
\end{equation} 
A brief outline of the covariant description of isotropic media can be found in \cite{demcsak2021response}, a more detailed description can be found in \cite{melrose2008quantum}.
\\
\\
Quantum plasmadynamics generalises quantum electrodynamics by replacing the vacuum electron propagator with a statistically averaged propagator describing the electron gas. Taking an average over a statistical distribution of electrons and positrons involves occupation numbers $n^{\epsilon}_{s}(\vec{p})$ describing electrons ($\epsilon = + 1$), and positrons ($\epsilon = - 1$), where $s=\pm1$ is the spin eigenvalue. 
\\
\\
Melrose and Parle \citep{melrose1983quantum} derived the general response of an arbitrary electron gas. It takes the form:
\begin{eqnarray}
\Pi^{\mu \nu} (k) =&& -e^{2} \sum_{\epsilon, \epsilon', s, s'} \int \frac{d^{3}\vec{p}}{(2\pi)^{3}} \int \frac{d^{3}\vec{p}\;'}{(2\pi)^{3}}(2\pi)^{3} \nonumber \\  &&\times \delta^{3} (\epsilon' \vec{p}\;' - \epsilon \vec{p} + \vec{k} ) \frac{\epsilon n_{s}^{\epsilon}(\vec{p}\;) - \epsilon' n_{s'}^{\epsilon'}(\vec{p}\;')}{\omega - \epsilon \varepsilon + \epsilon' \varepsilon' + i0}  \nonumber \\
&& \times [\Gamma _{s'\; s}^{\epsilon' \; \epsilon}(\vec{p\hspace{0.5mm}}\hspace{0.5mm}', \vec{p\hspace{0.5mm}}, \vec{k\hspace{0.5mm}} )]^{\mu}
[\Gamma _{s'\; s}^{\epsilon' \; \epsilon}(\vec{p\hspace{0.5mm}}\hspace{0.5mm}', \vec{p\hspace{0.5mm}}, \vec{k\hspace{0.5mm}} )]^{*\nu},
\label{4378}
\end{eqnarray}
with $\varepsilon = \sqrt{\vec{p}\;^{2} + m^{2}}$, and $\varepsilon' = \sqrt{\vec{p}^{\; \prime \hspace{0.5mm} 2} + m^{2}}$. The vertex functions corresponding to the linear response tensor take the form:
\begin{equation}
[\Gamma _{s'\; s}^{\epsilon' \; \epsilon}(\vec{p\hspace{0.5mm}}\hspace{0.5mm}', \vec{p\hspace{0.5mm}}, \vec{k\hspace{0.5mm}} )]^{\mu} = \frac{\bar{u}_{s'}^{\epsilon'}(\epsilon' \vec{p}\;') \gamma^{\mu} u_{s}^{\epsilon}(\epsilon \vec{p}\;)}{\sqrt{4\varepsilon\varepsilon'}},
\label{vertex11}
\end{equation}
where $\gamma^{\mu}$ are the Dirac matrices, and $u_{s}^{\epsilon}$, $\bar{u}_{s'}^{\epsilon'}$ are the electron and positron eigenfunctions respectively.
\\
\\
Occupation numbers may be separated into spin-specific, and spin-averaged contributions. One can explicitly identify the choice of spin eigenvalue for a given $\epsilon$ by writing:
\begin{equation}
n^{\epsilon} (\vec{p}) = \frac{1}{2} [n_{+}^{\epsilon} (\vec{p}\;) + n_{-}^{\epsilon} (\vec{p}\;)],
\end{equation}
\begin{equation}
\Delta n^{\epsilon} (\vec{p}) = \frac{1}{2} [n_{+}^{\epsilon} (\vec{p}\;) - n_{-}^{\epsilon} (\vec{p}\;)].
\end{equation}

\noindent Therefore the linear response tensor (Eq. \ref{4378}) can be written as $\Pi^{\mu \nu}(k)= \Pi^{\mu \nu}_{\textrm{in}}(k) +\Pi^{\mu \nu}_{\textrm{sd}}(k)$, where $\Pi^{\mu \nu}_{\textrm{in}}(k)$ is the spin-independent part, and   $\Pi^{\mu \nu}_{\textrm{sd}}(k)$ is the spin-dependent part. This follows from observing that we have in Eq. \ref{4378}:
\begin{eqnarray}
&&\sum_{s, s'} n_{s}^{\epsilon} (\vec{p}\;) [\Gamma _{s'\; s}^{\epsilon' \; \epsilon}(\vec{p\hspace{0.5mm}}\hspace{0.5mm}', \vec{p\hspace{0.5mm}}, \vec{k\hspace{0.5mm}} )]^{\mu} [\Gamma _{s'\; s}^{\epsilon' \; \epsilon}(\vec{p\hspace{0.5mm}}\hspace{0.5mm}', \vec{p\hspace{0.5mm}}, \vec{k\hspace{0.5mm}} )]^{*\nu} \nonumber \\
&&= n^{\epsilon}(\vec{p}\;)\mathcal{A} + \Delta n^{\epsilon} (\vec{p}\;) \mathcal{B},
\end{eqnarray}

\begin{eqnarray}
&&\sum_{s, s'}  n_{s'}^{\epsilon'} (\vec{p}\;') [\Gamma _{s'\; s}^{\epsilon' \; \epsilon}(\vec{p\hspace{0.5mm}}\hspace{0.5mm}', \vec{p\hspace{0.5mm}}, \vec{k\hspace{0.5mm}} )]^{\mu} [\Gamma _{s'\; s}^{\epsilon' \; \epsilon}(\vec{p\hspace{0.5mm}}\hspace{0.5mm}', \vec{p\hspace{0.5mm}}, \vec{k\hspace{0.5mm}} )]^{*\nu} \nonumber \\
&&= n^{\epsilon'}(\vec{p}\;')\mathcal{A} + \Delta n^{\epsilon'} (\vec{p}\;') \mathcal{C},
\end{eqnarray}
with
\begin{align}
\mathcal{A} &= \sum_{s, s'} \;[\Gamma _{s'\; s}^{\epsilon' \; \epsilon}(\vec{p\hspace{0.5mm}}\hspace{0.5mm}', \vec{p\hspace{0.5mm}}, \vec{k\hspace{0.5mm}} )]^{\mu}
[\Gamma _{s'\; s}^{\epsilon' \; \epsilon}(\vec{p\hspace{0.5mm}}\hspace{0.5mm}', \vec{p\hspace{0.5mm}}, \vec{k\hspace{0.5mm}} )]^{*\nu},\\
\mathcal{B} &= \sum_{s, s'} \; s \;[\Gamma _{s'\; s}^{\epsilon' \; \epsilon}(\vec{p\hspace{0.5mm}}\hspace{0.5mm}', \vec{p\hspace{0.5mm}}, \vec{k\hspace{0.5mm}} )]^{\mu}
[\Gamma _{s'\; s}^{\epsilon' \; \epsilon}(\vec{p\hspace{0.5mm}}\hspace{0.5mm}', \vec{p\hspace{0.5mm}}, \vec{k\hspace{0.5mm}} )]^{*\nu},\label{whereB}\\
\mathcal{C} &= \sum_{s, s'} \; s' \;[\Gamma _{s'\; s}^{\epsilon' \; \epsilon}(\vec{p\hspace{0.5mm}}\hspace{0.5mm}', \vec{p\hspace{0.5mm}}, \vec{k\hspace{0.5mm}} )]^{\mu}
[\Gamma _{s'\; s}^{\epsilon' \; \epsilon}(\vec{p\hspace{0.5mm}}\hspace{0.5mm}', \vec{p\hspace{0.5mm}}, \vec{k\hspace{0.5mm}} )]^{*\nu}. \label{whereC}
\end{align}
\\
The spin-independent part of the response tensor can be calculated by  evaluating a trace involving gamma matrices \cite{demcsak2021response, melrose2003optical} of the form:
\begin{equation}
F^{\mu \nu} (K, K') = \frac{1}{4} \textrm{Tr} [\gamma^{\mu}(\slashed{K}+m) \gamma^{\nu}(\slashed{K}' +m)],
\end{equation}
where $F^{\mu \nu}(\epsilon \tilde{p}, \epsilon' \tilde{p}') = \epsilon \varepsilon \epsilon' \varepsilon' \mathcal{A}$, and $\tilde{p} = [\varepsilon, \vec{p}\hspace{0.5mm}]$, $\tilde{p}\;' = [\varepsilon ', \vec{p}\; ']$. The slash notation is where 4-vectors are contracted with $\gamma$-matrices: $\slashed{K} = K_{\mu} \gamma^{\mu}$.
\\
\\
The spin-dependent part of the response tensor has two contributions:
\begin{eqnarray}
\Pi_{\textrm{sd}}^{\mu\nu}(k) &&= -e^{2} \sum_{\epsilon, \epsilon'} \int \frac{d^{3} \vec{p}}{(2\pi)^{3}} \frac{\epsilon \Delta n^{\epsilon}(\vec{p})}{\omega - \epsilon\varepsilon + \epsilon' \varepsilon' + i0} \; \mathcal{B} \nonumber \\
&& +e^{2} \sum_{\epsilon, \epsilon'} \int \frac{d^{3} \vec{p}\hspace{0.5mm}'}{(2\pi)^{3}} \frac{\epsilon' \Delta n^{\epsilon'}(\vec{p}\hspace{0.5mm}')}{\omega - \epsilon\varepsilon + \epsilon' \varepsilon' + i0} \;\mathcal{C}.
\label{sdtensor}
\end{eqnarray}

\section{Response of an electric moment dependent electron gas} \label{2133}
In this section we calculate Eq. \ref{whereB} and Eq. \ref{whereC}. These are the first and second contributions to the spin-dependent linear response tensor $\Pi^{\mu \nu}_{\textrm{sd}}(k)$ respectively. It should be noted that both $\mathcal{B}$ and $\mathcal{C}$  depend explicitly on the choice of spin operator. The nonzero components of $\mathcal{B}$ and $\mathcal{C}$ give the electric moment-dependent part of the response tensor.

\subsection{Electric moment eigenfunction}
The temporal evolution of the electric moment operator $\hat{\vec{\mathbb{D}}}$ is governed by ${d \hat{\vec{\mathbb{D}}}}/{dt} = i[\hat{H}, \hat{\vec{\mathbb{D}}}]$, where $\hat{H}$ is the Dirac Hamiltonian. In the presence of an electromagnetic field this becomes $ {d \hat{\vec{\mathbb{D}}}}/{dt} = ie \vec{\gamma} \times \vec{B} + e\gamma^{0} \vec{\sigma} \times \vec{E}$, where $\vec{\sigma}$ are the Pauli matrices in the $4\times 4$ Dirac spin space. It follows that $\hat{\mathbb{D}}$ is a constant of the motion if an electrostatic field is present along the $z$-axis. The $z$-component of the electric moment operator takes the following form:
\begin{equation}
\hat{\mathbb{D}}_{z}=
\left(
\begin{array}{cccc}
0 & p_{\perp}e^{-i\phi} & 0 & 0\\
-p_{\perp} e^{i\phi} & 0 & 0 & 0\\
0 & 0 & 0 & -p_{\perp} e^{-i\phi}\\
0 & 0 & p_{\perp} e^{i\phi} & 0\\
\end{array}
\right),
\label{mmoment0}
\end{equation}

\noindent and its eigenvalues are $s p_{\perp}$, where $s=\pm1$. The momentum is expressed using spherical polar coordinates $\vec{p} = (|\vec{p}\hspace{0.5mm}| \sin{\theta} \cos{\phi}, |\vec{p}\hspace{0.5mm}| \sin{\theta} \sin{\phi}, |\vec{p}\hspace{0.5mm}| \cos{\theta})$, where $p_{\perp} = |\vec{p}\hspace{0.5mm}| \sin{\theta}$ and $p_{z} = |\vec{p}\hspace{0.5mm}| \cos{\theta}$.
\\
\\
The spin operator $\hat{\mathbb{D}}_{z}$ is a relativistically acceptable spin operator \cite{sokolov1966synchrotron, sokolov1986radiation, fradkin1961electron}, since it commutes with the Dirac Hamiltonian which ensures the eigenvalues of $\hat{\mathbb{D}}_{z}$ are constants of the motion. The polarization states correspond to the eigenfunctions of the component of the electric-moment operator $\hat{\vec{\mathbb{D}}}$ along the direction of the electric field $\vec{E}$ used to separate the up and down spin states in creating the plasma.

\begin{widetext}
 We assume plane-wave solutions of the form $\exp{[-i\epsilon (\varepsilon t - \vec{p} \cdot \vec{x})]}$, where $\hat{\vec{p}} = -i\partial / \partial\vec{x}$ in the coordinate representation. A specific choice of simultaneous eigenfunctions of both $\hat{\mathbb{D}}_{z}$, and the Dirac Hamiltonian is:
\begin{equation}
u_{s}^{\epsilon}(\epsilon \vec{p} \hspace{0.7mm})=
\frac{1}{\sqrt{4 \varepsilon}}
\left(
\begin{array}{c}
\sqrt{\varepsilon + \epsilon m} \;\; e^{-i(\epsilon s \theta + \phi)/2} \\
i\epsilon s \sqrt{\varepsilon + \epsilon m} \;\;e^{-i(\epsilon s \theta - \phi)/2} \\
\sqrt{\varepsilon - \epsilon m}\;\;e^{i(\epsilon s \theta - \phi)/2}\\
-i\epsilon s \sqrt{\varepsilon - \epsilon m}\;\; e^{i(\epsilon s \theta + \phi)/2} \\
\end{array}
\right).
\label{mmoment1}
\end{equation}

\noindent The vertex function (Eq. \ref{vertex11}) for the electric moment eigenfunction (Eq. \ref{mmoment1}) is given by 
\begin{equation}
[\Gamma_{s, s'}^{\epsilon, \epsilon'} (\vec{p}\;', \vec{p}\;)]^{\mu}=
\frac{1}{\sqrt{16 \varepsilon \varepsilon'}}\\
\left(
\begin{array}{cccc}
d_{+}' d_{+} e^{i(c_{+}' - c_{+})/2} + s s' \epsilon \epsilon' d_{+}' d_{+} e^{i(c_{-}' - c_{-})/2} + d_{-}'d_{-} e^{i(-c_{-}' + c_{-})/2} + s s' \epsilon \epsilon' d_{-}' d_{-} e^{i(-c_{+}' + c_{+})/2} \\
-is\epsilon d_{+}' d_{-} e^{i(c_{+}' + c_{+})/2} - is'\epsilon' d_{+}' d_{-} e^{i(c_{-}' + c_{-})/2} + is\epsilon d_{-}' d_{+} e^{-i(c_{-}' + c_{-})/2} + is' \epsilon' d_{-}' d_{+} e^{-i(c_{+}' + c_{+})/2} \\
-s\epsilon d_{+}' d_{-} e^{i (c_{+}' + c_{+})/2} + s'\epsilon' d_{+}' d_{-} e^{i(c_{-}' + c_{-})/2} + s\epsilon d_{-}' d_{+} e^{-i(c_{-}' + c_{-})/2} - s' \epsilon' d_{-}' d_{+} e^{-i(c_{+}' + c_{+})/2}\\
d_{+}' d_{-} e^{i(c_{+}' + c_{-})/2} + s s' \epsilon \epsilon' d_{+}' d_{-} e^{i(c_{-}' + c_{+})/2} + d_{-}' d_{+} e^{-i(c_{-}' + c_{+})/2} + s s'\epsilon \epsilon' d_{-}' d_{+} e^{-i(c_{+}' + c_{-})/2}\\
\end{array}
\right),
\label{vertexmagmoment}
\end{equation}
where
$ c_{\pm} = \epsilon s \theta \pm \phi$, 
$d_{\pm} = \sqrt{\varepsilon \pm \epsilon m} $,
$ c_{\pm}' = \epsilon' s' \theta' \pm \phi'$, 
$d_{\pm}' = \sqrt{\varepsilon' \pm \epsilon' m} $, $d_{+}d_{-} = |\vec{p}\hspace{0.5mm}|$, $d'_{+}d'_{-} = |\vec{p}\hspace{0.5mm}'|$.
\\
\\
Direct evaluation of Eq. \ref{sdtensor} recovers the electric-moment dependent part of the response tensor.

 \begin{equation}
\Pi_{\textrm{sd}}^{\mu \nu}=
i m e^{2} k^{2}  \sum_{\epsilon} \int \frac{d^{3} \vec{p}}{(2\pi)^{3}}\frac{\epsilon \Delta n^{\epsilon}(\vec{p})}{(pk)^{2} - ({k^{2}}/{2})^{2}} \;
b^{\mu \nu}(k,p),
\nonumber 
\end{equation}

 \noindent where 

\begin{eqnarray}
b^{\mu \nu} (k,p)=
\frac{1}{\varepsilon p_{\perp}}\left(
\begin{array}{cccc}
0\;\;\;\;& -p_{x}k_{z} \;\;\;\;& -p_{y}k_{z} \;\;\;\;& p_{x}k_{x} + p_{y}k_{y}\;\;\; \\
p_{x}k_{z} & 0 & 0 & \omega p_{x} \\
p_{y}k_{z} & 0 & 0 & \omega p_{y} \\
-p_{x}k_{x} - p_{y}k_{y} & -\omega p_{x} &-\omega p_{y} &0 \\
\end{array}
\right),
\label{spinmagresponse2}
\end{eqnarray}

 \noindent with $pk = \omega \varepsilon - \vec{p} \cdot \vec{k}$, and $k^{2} = \omega^{2} - |\vec{k}\hspace{0.5mm}|^{2}$. A sample derivation  of Eq. \ref{spinmagresponse2} can be found in the appendix. The wavefunction $u_{s}^{\epsilon}(\epsilon \vec{p} \hspace{0.7mm})$ and $\bar{u}_{s}^{\epsilon}(\epsilon \vec{p} \hspace{0.7mm})$ are simultaneous eigenfunctions of both $\hat{\mathbb{D}}_{z}$ and $\hat{H}
 $. By inspection of Eq. \ref{spinmagresponse2} it is clear that the gauge invariance and charge continuity conditions (Eq. \ref{gaugecharge}) are satisfied. 
 \\
 \\
By inspection of Eq. \ref{spinmagresponse2} one can see that the linear response for an arbitrary electric moment dependent isotropic (in momentum) quantum plasma is zero. It is nonzero if there is a streaming motion perpendicular to the axis, such that $\langle p_x\rangle\ne0$ or $\langle p_y\rangle\ne0$, however this is only true if the electron and positron distributions are not identical.
When $\langle p_y\rangle\ne0$ the response tensor agrees with that of a plasma having a nonzero, cross-field current.


\section{Relativistic Quantum Correction to a Streaming Cold Plasma}
The dispersion equation for a background plasma with a spin correction factor takes the form: 
\begin{equation}
\det[\lambda_{i j}] = \det[n^{2}(\kappa_{i}\kappa_{j} - \delta_{ij}) + (\Pi_{i j} + K_{i j})] =0,
\label{disper445}
\end{equation}
 where $n=kc/\omega$ is the refractive index (in natural units $c \rightarrow 1$), $\Pi_{i j}$ is the spin-dependent correction, $\vec{\kappa}$ is the unit vector of $\vec{k}$, and $K_{i j}$ is the dielectric tensor describing the background plasma. Properties of the wave modes can then be identified from the dispersion equation.
\\
\\
From the spin corrected 4-response $\Pi_{\mu \nu}$ (Eq. \ref{spinmagresponse2}), one can extract the spin corrected 3-response $\Pi_{ij}$ by omitting the first row and first column. The 3-response describing a cold plasma takes the form:
\begin{equation}
K_{ij} = \delta_{ij} - \frac{\omega_{p}^{2}}{\omega^{2}} \left[\delta_{ij} + \frac{k_{i}v_{j} + k_{j}v_{i}}{\omega - \vec{k}\cdot \vec{v}} + \frac{v_{i}v_{j} \left(|\vec{k}|^{2} - \frac{\omega^{2}}{c^{2}} \right)}{(\omega - \vec{k}\cdot\vec{v})^{2}} \right].
\label{cp33}
\end{equation}
When a cold plasma is streaming in the y-direction, one has $v_{y}\neq0$ and $v_{x} = v_{z}=0$. Eq. \ref{cp33} has the following symmetry: $K_{13} = K_{31}, K_{11}=K_{33}$ and $K_{12} = K_{21}$. By inspection of Eq. \ref{spinmagresponse2}, there are two non-zero spin-correction contributions applicable to a plasma streaming in the y-direction ($\mathbb{B}_{32} = - \mathbb{B}_{23}$): 
$$\mathbb{B}_{23} = \frac{i \omega \langle p_{y} \rangle}{\varepsilon p_{\perp}} \mathbb{H}, \hspace{10mm} \textrm{where} \hspace{10mm} \mathbb{H} = \frac{me^{2}k^{2}}{(2\pi)^{3}} \sum_{\epsilon} \frac{\epsilon \Delta n^{\epsilon}(\vec{p}\hspace{0.5mm})}{(pk)^{2} - (k^{2}/2)^{2}}.$$
The momentum is expressed using spherical polar coordinates in the electric moment operator $\mathbb{D}_{z}$. Thus one must pick $\kappa_{i} = (\sin \theta \cos\phi, \sin \theta \sin \phi, \cos \theta)$. We can now directly evaluate the dispersion equation $\det{\lambda_{ij}}=0$ using:

 \begin{eqnarray}
\lambda_{ij} =   \left(
\begin{array}{ccc}
n^{2}\sin^{2}\theta \cos^{2}\phi - n^{2} + K_{11}&n^{2}\sin^{2}\theta \sin {\phi} \cos\phi + K_{12}&n^{2}\sin\theta \cos\theta \cos\phi\\
n^{2}\sin^{2}\theta\sin\phi\cos\phi + K_{12}&n^{2}\sin^{2}\theta\sin^{2}\phi - n^{2} + K_{22}&n^{2}\sin\theta\cos\theta\sin\phi + K_{23} + \mathbb{B}_{23}\\
n^{2}\sin\theta\cos\theta\cos\phi&n^{2}\sin\theta\cos\theta\sin\phi + K_{23} - \mathbb{B}_{23}&n^{2}\cos^{2}\theta - n^{2} + K_{11}\\
\end{array}
\right)
\hspace{4mm}
\label{dispersion1}
\end{eqnarray}

\noindent Eq. \ref{dispersion1} takes the form $\det[\lambda_{ij}] = An^{4} + B{n^2} + C$, where $A=\kappa_i\kappa_j K_{ij}$, $B=AK_{ss}-(K_{is}K_{sj}-\mathbb{B}_{is} \mathbb{B}_{sj})\kappa_i\kappa_j$, and $C=\det[K_{ij}+\mathbb{B}_{ij}]$.
\\
\\
Detailed properties of the wave modes for an electric moment dependent relativistic quantum plasma can be investigated by interpreting the dispersion relation calculated using Eq. \ref{dispersion1}. To simplify this calculation we set $k_{i} \rightarrow \omega/c$, and $n \rightarrow 1 $ in the term $(\Pi_{i j} + K_{i j})$ of Eq. \ref{disper445}. The dispersion equation can then be solved for n, from which the wave modes can be identified. In the case of an electric moment dependent quantum plasma, further insight follows from investigating the polarisation vectors. One can find the components of the polarisation vector by normalising the components in any column of the matrix of cofactors of Eq. \ref{dispersion1}. One finds that the $x$ and $y$-components of the first column are real, and the $z$-component is complex. This corresponds to a polarisation vector whose transverse component is elliptical, this behaviour is different to both the optically active and gyrotropic cases (which respectively correspond to a helicity polarised, and a magnetic moment polarised relativistic quantum plasma).

\section{Discussion and Conclusions}
Spin dependent effects in relativistic quantum plasmas must be investigated using an approach that ensures that the spin operator in question commutes with the Dirac Hamiltonian. If this condition is not satisfied, physical interpretations cannot be made. In particular, it has been established by Demcsak and Melrose \citep{demcsak2021response} that the response of a spin-dependent electron gas depends on the particular spin operator in question. The result derived in this Letter shows that quantum plasmas that are spin polarised by the helicity, magnetic moment, and electric moment operators all have distinct spin-dependent responses. In particular, they are optically active \cite{melrose2003optical}, gyrotropic \cite{demcsak2021response}, and non-spatially dispersive respectively. \\
\\
Differences between the spin-dependent contributions of response tensors corresponding to distinct spin operators is intrinsically quantum mechanical, and therefore a small effect. However these differences are in principle observable. The operator approach of quantum plasmadynamics allows one to readily include relativistically acceptable spin-operators to quantum plasmas. This is in contrast to the Wigner function approaches to quantum plasmas, where the distinction between spin operators is absent or obscure - obscuring potentially observable consequences of the difference between spin operators. 

  \begin{acknowledgments}

\end{acknowledgments}
  \end{widetext}

\appendix*
\begin{widetext}
\section{Derivation of $b^{03}$}
The derivation of the 13-component of Eq. \ref{spinmagresponse2} proceeds as follows. We start with taking the product of the vertex functions corresponding to the tensor entry to be computed.
\begin{eqnarray*}
[\Gamma]^{1}[\Gamma]^{*2}=\frac{1}{16 \varepsilon  \varepsilon'} &&\left[ 
d_{+}' d_{+} e^{i(c_{+}' - c_{+})/2} + s s' \epsilon \epsilon' d_{+}' d_{+} e^{i(c_{-}' - c_{-})/2} + d_{-}'d_{-} e^{i(-c_{-}' + c_{-})/2} + s s' \epsilon \epsilon' d_{-}' d_{-} e^{i(-c_{+}' + c_{+})/2}
\right] \\ 
&& \times \left[d_{+}' d_{-} e^{-i/2(c_{+}' + c_{-})} + s s' \epsilon \epsilon' d_{+}' d_{-} e^{-i/2(c_{-}' + c_{+})} + d_{-}' d_{+} e^{i/2(c_{-}' + c_{+})} + s s' \epsilon \epsilon' d_{-}' d_{+} e^{i/2(c_{+}' + c_{-})} \right],
\end{eqnarray*}
where
\begin{eqnarray}
\noindent \sum_{s, s'} s [\Gamma]^{0}[\Gamma]^{*3} = \frac{im}{2\varepsilon \varepsilon' p_{\perp}}\left[-\epsilon \epsilon' + p'_{x} p_{x} + p'_{y} p_{y}\right],
\end{eqnarray}
and
\begin{eqnarray}
\sum_{s, s'} s' [\Gamma]^{0}[\Gamma]^{*3} = \frac{im}{2\varepsilon \varepsilon' p'_{\perp}} \left[\epsilon \epsilon' - p'_{x}p_{x} - p'_{y}p_{y} \right].
\end{eqnarray}

The spin-dependent part of the response tensor is the sum of the two terms in Eq. \ref{sdtensor}

\begin{eqnarray}
\Pi^{03}_{\textrm{sd}} = -e^{2} \sum_{\epsilon, \epsilon'} \int \frac{d^{3} \vec{p}}{(2\pi)^{3}} \frac{\Delta n^{\epsilon}(\vec{p})}{\omega - \epsilon\varepsilon + \epsilon' \varepsilon' }  \frac{im}{2\varepsilon \varepsilon' p_{\perp}}\left[-\epsilon \epsilon' + p'_{x} p_{x} + p'_{y} p_{y}\right] \\
 +e^{2} \sum_{\epsilon, \epsilon'} \int  \frac{d^{3} \vec{p}\hspace{0.5mm}'}{(2\pi)^{3}} \frac{\Delta n^{\epsilon'}(\vec{p}\hspace{0.5mm}')}{\omega - \epsilon\varepsilon + \epsilon' \varepsilon' } \frac{im}{2\varepsilon \varepsilon' p'_{\perp}} \left[\epsilon \epsilon' - p'_{x}p_{x} - p'_{y}p_{y} \right].
\end{eqnarray}

\noindent The conservation of momentum relations $\epsilon' \vec{p}\hspace{0.5mm}' = \epsilon \vec{p} - \vec{k}$, and $\epsilon' \vec{\varepsilon}\hspace{0.5mm}' = \epsilon \vec{\varepsilon} - \vec{\omega}$ can be used to evaluate both contributions to the response tensor as follows. In the first term the primed quantities are replaced with unprimed quantities: $p' = \epsilon \epsilon' p - \epsilon' k$. In the second term the unprimed quantities are replaced with primed quantities: $p = \epsilon \epsilon' p' + \epsilon k$. The result is:

\begin{eqnarray}
\Pi^{03}_{\textrm{sd}} = -e^{2}&&\sum_{\epsilon, \epsilon'} \int  \frac{d^{3} \vec{p}}{(2\pi)^{3}} \frac{ \Delta n^{\epsilon}(\vec{p})}{\omega - \epsilon\varepsilon + \epsilon' \varepsilon' } \frac{im}{2\varepsilon \varepsilon' p_{\perp}}[-\epsilon' + \epsilon'p_{x}^{2} - \epsilon \epsilon'p_{x}k_{x} + \epsilon'p_{y}^{2} - \epsilon \epsilon' p_{y}k_{y}]
 \\
 && +e^{2} \sum_{\epsilon, \epsilon'} \int  \frac{d^{3} \vec{p}\hspace{0.5mm}'}{(2\pi)^{3}} \frac{ \Delta n^{\epsilon'}(\vec{p}\hspace{0.5mm}')}{\omega - \epsilon\varepsilon + \epsilon' \varepsilon'}
 \frac{im}{2 \varepsilon \varepsilon' p'_{\perp}} [\epsilon - \epsilon p_{x}'^{2} - \epsilon \epsilon' p'_{x} k_{x} - \epsilon p_{y}'^{2} - \epsilon \epsilon' p'_{y} k_{y}].
\end{eqnarray}

\noindent The next step is to evaluate the sums over $\epsilon' = \pm$ in the first term, and sums over $\epsilon = \pm$ in the second term. This requires the use of the following identities:
\begin{equation}
\left[\frac{1}{\omega - \epsilon \varepsilon + \varepsilon'} - \frac{1}{\omega - \epsilon \varepsilon - \varepsilon'} \right] = \frac{-2\varepsilon'}{-2\epsilon pk + k^{2}},
\end{equation}
\begin{equation}
\left[\frac{1}{\omega - \varepsilon + \epsilon' \varepsilon'} - \frac{1}{\omega +  \varepsilon + \epsilon' \varepsilon'} \right] = \frac{2\varepsilon}{2\epsilon' p' k + k^{2}}.
\end{equation}
The result is:
\begin{align}
\Pi^{03}_{\textrm{sd}} =
&-ime^{2} \sum_{\epsilon} \int \frac{d^{3}\vec{p}}{(2\pi)^{3}} \frac{ \epsilon \Delta n^{\epsilon}(\vec{p})}{\varepsilon p_{\perp}} [-p_{x}k_{x} - p_{y}k_{y}] \left(\frac{-1}{-2\epsilon pk + k^{2}} \right)  \\
&+ime^{2} \sum_{\epsilon'} \int \frac{d^{3}\vec{p\hspace{0.5mm}'}}{(2\pi)^{3}} \frac{ \epsilon ' \Delta n^{\epsilon'}(\vec{p} \hspace{0.5mm}')}{\varepsilon' p'_{\perp}} [-p'_{x} k_{x} - p'_{y}k_{y}] \left(\frac{1}{2\epsilon' p' k + k^{2}} \right) .
\end{align}

\noindent The next step is to swap the dummy variables present in the second contribution to the response tensor as follows: $p'_{\perp}, p', \varepsilon', \epsilon' \rightarrow p_{\perp}, p, \varepsilon, \epsilon$.  The two contributions to the response tensor are now combined into a single expression:
\begin{align*}
\Pi^{03}_{\textrm{sd}} =
 \; -ime^{2} \sum_{\epsilon} \int & \frac{d^{3}\vec{p}}{(2\pi)^{3}}\; \epsilon \Delta n^{\epsilon}(\vec{p}) \left(\frac{p_{x}k_{x} + p_{y} k_{y}}{\varepsilon p_{\perp}} \right) \left(\frac{1}{-2\epsilon pk + k^{2}} - \frac{1}{2\epsilon pk + k^{2}} \right) \\
= \;  -ime^{2} \sum_{\epsilon}& \int  \frac{d^{3}\vec{p}}{(2\pi)^{3}} \epsilon \Delta n^{\epsilon}(\vec{p}) \left(\frac{p_{x}k_{x} + p_{y} k_{y}}{\varepsilon p_{\perp}} \right) \left[\frac{k^{2}}{(pk)^{2} - (k^{2}/2)^{2}} \right],
 \\
= \;  ime^{2}k^{2} \sum_{\epsilon}& \int  \frac{d^{3}\vec{p}}{(2\pi)^{3}} \frac{\epsilon \Delta n^{\epsilon}(\vec{p})}{(pk)^{2} - (k^{2}/2)^{2}} \left(\frac{p_{x}k_{x} + p_{y} k_{y}}{\varepsilon p_{\perp}} \right).
\end{align*}

By inspection of the above one recovers the 03-component appearing in Eq. \ref{spinmagresponse2}. The derivations of the other components of the response tensor follow in a similar way.

\end{widetext}

\nocite{*}

\bibliography{elecbib}

\providecommand{\noopsort}[1]{}\providecommand{\singleletter}[1]{#1}%
\begin{thebibliography}{25}%
\makeatletter
\providecommand \@ifxundefined [1]{%
 \@ifx{#1\undefined}
}%
\providecommand \@ifnum [1]{%
 \ifnum #1\expandafter \@firstoftwo
 \else \expandafter \@secondoftwo
 \fi
}%
\providecommand \@ifx [1]{%
 \ifx #1\expandafter \@firstoftwo
 \else \expandafter \@secondoftwo
 \fi
}%
\providecommand \natexlab [1]{#1}%
\providecommand \enquote  [1]{``#1''}%
\providecommand \bibnamefont  [1]{#1}%
\providecommand \bibfnamefont [1]{#1}%
\providecommand \citenamefont [1]{#1}%
\providecommand \href@noop [0]{\@secondoftwo}%
\providecommand \href [0]{\begingroup \@sanitize@url \@href}%
\providecommand \@href[1]{\@@startlink{#1}\@@href}%
\providecommand \@@href[1]{\endgroup#1\@@endlink}%
\providecommand \@sanitize@url [0]{\catcode `\\12\catcode `\$12\catcode
  `\&12\catcode `\#12\catcode `\^12\catcode `\_12\catcode `\%12\relax}%
\providecommand \@@startlink[1]{}%
\providecommand \@@endlink[0]{}%
\providecommand \url  [0]{\begingroup\@sanitize@url \@url }%
\providecommand \@url [1]{\endgroup\@href {#1}{\urlprefix }}%
\providecommand \urlprefix  [0]{URL }%
\providecommand \Eprint [0]{\href }%
\providecommand \doibase [0]{https://doi.org/}%
\providecommand \selectlanguage [0]{\@gobble}%
\providecommand \bibinfo  [0]{\@secondoftwo}%
\providecommand \bibfield  [0]{\@secondoftwo}%
\providecommand \translation [1]{[#1]}%
\providecommand \BibitemOpen [0]{}%
\providecommand \bibitemStop [0]{}%
\providecommand \bibitemNoStop [0]{.\EOS\space}%
\providecommand \EOS [0]{\spacefactor3000\relax}%
\providecommand \BibitemShut  [1]{\csname bibitem#1\endcsname}%
\let\auto@bib@innerbib\@empty
\bibitem [{\citenamefont {Haas}(2011)}]{haas2011quantum}%
  \BibitemOpen
  \bibfield  {author} {\bibinfo {author} {\bibfnamefont {F.}~\bibnamefont
  {Haas}},\ }\href@noop {} {\emph {\bibinfo {title} {Quantum plasmas: An
  hydrodynamic approach}}},\ Vol.~\bibinfo {volume} {65}\ (\bibinfo
  {publisher} {Springer Science \& Business Media},\ \bibinfo {year}
  {2011})\BibitemShut {NoStop}%
\bibitem [{\citenamefont {Bonitz}(2016)}]{bonitz2016quantum}%
  \BibitemOpen
  \bibfield  {author} {\bibinfo {author} {\bibfnamefont {M.}~\bibnamefont
  {Bonitz}},\ }\href@noop {} {\emph {\bibinfo {title} {Quantum kinetic
  theory}}},\ Vol.\ \bibinfo {volume} {412}\ (\bibinfo  {publisher}
  {Springer},\ \bibinfo {year} {2016})\BibitemShut {NoStop}%
\bibitem [{\citenamefont {Bialynicki-Birula}\ \emph {et~al.}(1991)\citenamefont
  {Bialynicki-Birula}, \citenamefont {Gornicki},\ and\ \citenamefont
  {Rafelski}}]{bialynicki1991phase}%
  \BibitemOpen
  \bibfield  {author} {\bibinfo {author} {\bibfnamefont {I.}~\bibnamefont
  {Bialynicki-Birula}}, \bibinfo {author} {\bibfnamefont {P.}~\bibnamefont
  {Gornicki}},\ and\ \bibinfo {author} {\bibfnamefont {J.}~\bibnamefont
  {Rafelski}},\ }\bibfield  {title} {\bibinfo {title} {Phase-space structure of
  the dirac vacuum},\ }\href@noop {} {\bibfield  {journal} {\bibinfo  {journal}
  {Physical Review D}\ }\textbf {\bibinfo {volume} {44}},\ \bibinfo {pages}
  {1825} (\bibinfo {year} {1991})}\BibitemShut {NoStop}%
\bibitem [{\citenamefont {Vasak}\ \emph {et~al.}(1987)\citenamefont {Vasak},
  \citenamefont {Gyulassy},\ and\ \citenamefont {Elze}}]{vasak1987quantum}%
  \BibitemOpen
  \bibfield  {author} {\bibinfo {author} {\bibfnamefont {D.}~\bibnamefont
  {Vasak}}, \bibinfo {author} {\bibfnamefont {M.}~\bibnamefont {Gyulassy}},\
  and\ \bibinfo {author} {\bibfnamefont {H.-T.}\ \bibnamefont {Elze}},\
  }\bibfield  {title} {\bibinfo {title} {Quantum transport theory for abelian
  plasmas},\ }\href@noop {} {\bibfield  {journal} {\bibinfo  {journal} {Annals
  of Physics}\ }\textbf {\bibinfo {volume} {173}},\ \bibinfo {pages} {462}
  (\bibinfo {year} {1987})}\BibitemShut {NoStop}%
\bibitem [{\citenamefont {Hebenstreit}\ \emph {et~al.}(2010)\citenamefont
  {Hebenstreit}, \citenamefont {Alkofer},\ and\ \citenamefont
  {Gies}}]{hebenstreit2010schwinger}%
  \BibitemOpen
  \bibfield  {author} {\bibinfo {author} {\bibfnamefont {F.}~\bibnamefont
  {Hebenstreit}}, \bibinfo {author} {\bibfnamefont {R.}~\bibnamefont
  {Alkofer}},\ and\ \bibinfo {author} {\bibfnamefont {H.}~\bibnamefont
  {Gies}},\ }\bibfield  {title} {\bibinfo {title} {Schwinger pair production in
  space-and time-dependent electric fields: Relating the wigner formalism to
  quantum kinetic theory},\ }\href@noop {} {\bibfield  {journal} {\bibinfo
  {journal} {Physical Review D}\ }\textbf {\bibinfo {volume} {82}},\ \bibinfo
  {pages} {105026} (\bibinfo {year} {2010})}\BibitemShut {NoStop}%
\bibitem [{\citenamefont {Wigner}(1997)}]{wigner1997quantum}%
  \BibitemOpen
  \bibfield  {author} {\bibinfo {author} {\bibfnamefont {E.~P.}\ \bibnamefont
  {Wigner}},\ }\bibfield  {title} {\bibinfo {title} {On the quantum correction
  for thermodynamic equilibrium},\ }in\ \href@noop {} {\emph {\bibinfo
  {booktitle} {Part I: Physical Chemistry. Part II: Solid State Physics}}}\
  (\bibinfo  {publisher} {Springer},\ \bibinfo {year} {1997})\ pp.\ \bibinfo
  {pages} {110--120}\BibitemShut {NoStop}%
\bibitem [{\citenamefont {Melrose}(2020)}]{melrose2020quantum}%
  \BibitemOpen
  \bibfield  {author} {\bibinfo {author} {\bibfnamefont {D.}~\bibnamefont
  {Melrose}},\ }\bibfield  {title} {\bibinfo {title} {Quantum kinetic theory
  for unmagnetized and magnetized plasmas},\ }\href@noop {} {\bibfield
  {journal} {\bibinfo  {journal} {Reviews of Modern Plasma Physics}\ }\textbf
  {\bibinfo {volume} {4}},\ \bibinfo {pages} {1} (\bibinfo {year}
  {2020})}\BibitemShut {NoStop}%
\bibitem [{\citenamefont {Demcsak}\ and\ \citenamefont
  {Melrose}(2021)}]{demcsak2021response}%
  \BibitemOpen
  \bibfield  {author} {\bibinfo {author} {\bibfnamefont {V.~M.}\ \bibnamefont
  {Demcsak}}\ and\ \bibinfo {author} {\bibfnamefont {D.~B.}\ \bibnamefont
  {Melrose}},\ }\bibfield  {title} {\bibinfo {title} {Response tensor for a
  spin-dependent electron gas: dependence on the choice of spin operator},\
  }\href@noop {} {\bibfield  {journal} {\bibinfo  {journal} {Physical Review
  E}\ }\textbf {\bibinfo {volume} {103}},\ \bibinfo {pages} {L061201} (\bibinfo
  {year} {2021})}\BibitemShut {NoStop}%
\bibitem [{\citenamefont {Melrose}\ and\ \citenamefont
  {Weise}(2003)}]{melrose2003optical}%
  \BibitemOpen
  \bibfield  {author} {\bibinfo {author} {\bibfnamefont {D.}~\bibnamefont
  {Melrose}}\ and\ \bibinfo {author} {\bibfnamefont {J.}~\bibnamefont
  {Weise}},\ }\bibfield  {title} {\bibinfo {title} {Optical activity in an
  isotropic gas of electrons with a preferred helicity},\ }\href@noop {}
  {\bibfield  {journal} {\bibinfo  {journal} {Physical Review E}\ }\textbf
  {\bibinfo {volume} {68}},\ \bibinfo {pages} {046404} (\bibinfo {year}
  {2003})}\BibitemShut {NoStop}%
\bibitem [{\citenamefont {Blinne}\ and\ \citenamefont
  {Gies}(2014)}]{blinne2014pair}%
  \BibitemOpen
  \bibfield  {author} {\bibinfo {author} {\bibfnamefont {A.}~\bibnamefont
  {Blinne}}\ and\ \bibinfo {author} {\bibfnamefont {H.}~\bibnamefont {Gies}},\
  }\bibfield  {title} {\bibinfo {title} {Pair production in rotating electric
  fields},\ }\href@noop {} {\bibfield  {journal} {\bibinfo  {journal} {Physical
  Review D}\ }\textbf {\bibinfo {volume} {89}},\ \bibinfo {pages} {085001}
  (\bibinfo {year} {2014})}\BibitemShut {NoStop}%
\bibitem [{\citenamefont {Blinne}\ and\ \citenamefont
  {Strobel}(2016)}]{blinne2016comparison}%
  \BibitemOpen
  \bibfield  {author} {\bibinfo {author} {\bibfnamefont {A.}~\bibnamefont
  {Blinne}}\ and\ \bibinfo {author} {\bibfnamefont {E.}~\bibnamefont
  {Strobel}},\ }\bibfield  {title} {\bibinfo {title} {Comparison of
  semiclassical and wigner function methods in pair production in rotating
  fields},\ }\href@noop {} {\bibfield  {journal} {\bibinfo  {journal} {Physical
  Review D}\ }\textbf {\bibinfo {volume} {93}},\ \bibinfo {pages} {025014}
  (\bibinfo {year} {2016})}\BibitemShut {NoStop}%
\bibitem [{\citenamefont {Melrose}(2008)}]{melrose2008quantum}%
  \BibitemOpen
  \bibfield  {author} {\bibinfo {author} {\bibfnamefont {D.}~\bibnamefont
  {Melrose}},\ }\href@noop {} {\emph {\bibinfo {title} {Quantum Plasmadynamics:
  unmagnetised plasmas}}},\ Vol.\ \bibinfo {volume} {735}\ (\bibinfo
  {publisher} {Springer},\ \bibinfo {year} {2008})\BibitemShut {NoStop}%
\bibitem [{\citenamefont {Melrose}\ and\ \citenamefont
  {Parle}(1983)}]{melrose1983quantum}%
  \BibitemOpen
  \bibfield  {author} {\bibinfo {author} {\bibfnamefont {D.}~\bibnamefont
  {Melrose}}\ and\ \bibinfo {author} {\bibfnamefont {A.}~\bibnamefont
  {Parle}},\ }\bibfield  {title} {\bibinfo {title} {Quantum electrodynamics in
  strong magnetic fields. iii. electron-photon interactions},\ }\href@noop {}
  {\bibfield  {journal} {\bibinfo  {journal} {Australian Journal of Physics}\
  }\textbf {\bibinfo {volume} {36}},\ \bibinfo {pages} {799} (\bibinfo {year}
  {1983})}\BibitemShut {NoStop}%
\bibitem [{\citenamefont {Sokolov}\ and\ \citenamefont
  {Ternov}(1966)}]{sokolov1966synchrotron}%
  \BibitemOpen
  \bibfield  {author} {\bibinfo {author} {\bibfnamefont {A.~A.}\ \bibnamefont
  {Sokolov}}\ and\ \bibinfo {author} {\bibfnamefont {I.~M.}\ \bibnamefont
  {Ternov}},\ }\bibfield  {title} {\bibinfo {title} {Synchrotron radiation},\
  }\href@noop {} {\bibfield  {journal} {\bibinfo  {journal} {Soviet Physics
  Journal}\ } (\bibinfo {year} {1966})}\BibitemShut {NoStop}%
\bibitem [{\citenamefont {Sokolov}\ and\ \citenamefont
  {Ternov}(1986)}]{sokolov1986radiation}%
  \BibitemOpen
  \bibfield  {author} {\bibinfo {author} {\bibfnamefont {A.~A.}\ \bibnamefont
  {Sokolov}}\ and\ \bibinfo {author} {\bibfnamefont {I.~M.}\ \bibnamefont
  {Ternov}},\ }\href@noop {} {\emph {\bibinfo {title} {Radiation from
  relativistic electrons}}}\ (\bibinfo  {publisher} {AIP},\ \bibinfo {year}
  {1986})\BibitemShut {NoStop}%
\bibitem [{\citenamefont {Fradkin}\ and\ \citenamefont
  {Good~Jr}(1961)}]{fradkin1961electron}%
  \BibitemOpen
  \bibfield  {author} {\bibinfo {author} {\bibfnamefont {D.}~\bibnamefont
  {Fradkin}}\ and\ \bibinfo {author} {\bibfnamefont {R.}~\bibnamefont
  {Good~Jr}},\ }\bibfield  {title} {\bibinfo {title} {Electron polarization
  operators},\ }\href@noop {} {\bibfield  {journal} {\bibinfo  {journal}
  {Reviews of Modern Physics}\ }\textbf {\bibinfo {volume} {33}},\ \bibinfo
  {pages} {343} (\bibinfo {year} {1961})}\BibitemShut {NoStop}%
\bibitem [{\citenamefont {Ekman}\ \emph {et~al.}(2017)\citenamefont {Ekman},
  \citenamefont {Asenjo},\ and\ \citenamefont
  {Zamanian}}]{ekman2017relativistic}%
  \BibitemOpen
  \bibfield  {author} {\bibinfo {author} {\bibfnamefont {R.}~\bibnamefont
  {Ekman}}, \bibinfo {author} {\bibfnamefont {F.}~\bibnamefont {Asenjo}},\ and\
  \bibinfo {author} {\bibfnamefont {J.}~\bibnamefont {Zamanian}},\ }\bibfield
  {title} {\bibinfo {title} {Relativistic kinetic equation for spin-1/2
  particles in the long-scale-length approximation},\ }\href@noop {} {\bibfield
   {journal} {\bibinfo  {journal} {Physical Review E}\ }\textbf {\bibinfo
  {volume} {96}},\ \bibinfo {pages} {023207} (\bibinfo {year}
  {2017})}\BibitemShut {NoStop}%
\bibitem [{\citenamefont {Melrose}(2012)}]{melrose2012quantum}%
  \BibitemOpen
  \bibfield  {author} {\bibinfo {author} {\bibfnamefont {D.}~\bibnamefont
  {Melrose}},\ }\href@noop {} {\emph {\bibinfo {title} {Quantum plasmadynamics:
  magnetized plasmas}}},\ Vol.\ \bibinfo {volume} {854}\ (\bibinfo  {publisher}
  {Springer},\ \bibinfo {year} {2012})\BibitemShut {NoStop}%
\bibitem [{\citenamefont {Silenko}(2009)}]{silenko2009foldy}%
  \BibitemOpen
  \bibfield  {author} {\bibinfo {author} {\bibfnamefont {A.~J.}\ \bibnamefont
  {Silenko}},\ }\bibfield  {title} {\bibinfo {title} {Foldy-wouthuysen
  transformation and semiclassical transition for relativistic quantum
  mechanics},\ }\href@noop {} {\bibfield  {journal} {\bibinfo  {journal} {arXiv
  preprint arXiv:0910.5155}\ } (\bibinfo {year} {2009})}\BibitemShut {NoStop}%
\bibitem [{\citenamefont {Melrose}\ and\ \citenamefont
  {Hardy}(1996)}]{melrose1996quantum}%
  \BibitemOpen
  \bibfield  {author} {\bibinfo {author} {\bibfnamefont {D.}~\bibnamefont
  {Melrose}}\ and\ \bibinfo {author} {\bibfnamefont {S.}~\bibnamefont
  {Hardy}},\ }\bibfield  {title} {\bibinfo {title} {Quantum plasmadynamics:
  role of the electron self-energy and the vertex correction},\ }\href@noop {}
  {\bibfield  {journal} {\bibinfo  {journal} {Journal of plasma physics}\
  }\textbf {\bibinfo {volume} {56}},\ \bibinfo {pages} {95} (\bibinfo {year}
  {1996})}\BibitemShut {NoStop}%
\bibitem [{\citenamefont {Ekman}\ \emph {et~al.}(2019)\citenamefont {Ekman},
  \citenamefont {Al-Naseri}, \citenamefont {Zamanian},\ and\ \citenamefont
  {Brodin}}]{ekman2019relativistic}%
  \BibitemOpen
  \bibfield  {author} {\bibinfo {author} {\bibfnamefont {R.}~\bibnamefont
  {Ekman}}, \bibinfo {author} {\bibfnamefont {H.}~\bibnamefont {Al-Naseri}},
  \bibinfo {author} {\bibfnamefont {J.}~\bibnamefont {Zamanian}},\ and\
  \bibinfo {author} {\bibfnamefont {G.}~\bibnamefont {Brodin}},\ }\bibfield
  {title} {\bibinfo {title} {Relativistic kinetic theory for spin-1/2
  particles: Conservation laws, thermodynamics, and linear waves},\ }\href@noop
  {} {\bibfield  {journal} {\bibinfo  {journal} {Physical Review E}\ }\textbf
  {\bibinfo {volume} {100}},\ \bibinfo {pages} {023201} (\bibinfo {year}
  {2019})}\BibitemShut {NoStop}%
\bibitem [{\citenamefont {Hurst}\ \emph {et~al.}(2017)\citenamefont {Hurst},
  \citenamefont {Hervieux},\ and\ \citenamefont {Manfredi}}]{hurst2017phase}%
  \BibitemOpen
  \bibfield  {author} {\bibinfo {author} {\bibfnamefont {J.}~\bibnamefont
  {Hurst}}, \bibinfo {author} {\bibfnamefont {P.-A.}\ \bibnamefont
  {Hervieux}},\ and\ \bibinfo {author} {\bibfnamefont {G.}~\bibnamefont
  {Manfredi}},\ }\bibfield  {title} {\bibinfo {title} {Phase-space methods for
  the spin dynamics in condensed matter systems},\ }\href@noop {} {\bibfield
  {journal} {\bibinfo  {journal} {Philosophical Transactions of the Royal
  Society A: Mathematical, Physical and Engineering Sciences}\ }\textbf
  {\bibinfo {volume} {375}},\ \bibinfo {pages} {20160199} (\bibinfo {year}
  {2017})}\BibitemShut {NoStop}%
\bibitem [{\citenamefont {Melrose}\ and\ \citenamefont
  {McPhedran}(2005)}]{melrose2005electromagnetic}%
  \BibitemOpen
  \bibfield  {author} {\bibinfo {author} {\bibfnamefont {D.~B.}\ \bibnamefont
  {Melrose}}\ and\ \bibinfo {author} {\bibfnamefont {R.~C.}\ \bibnamefont
  {McPhedran}},\ }\href@noop {} {\emph {\bibinfo {title} {Electromagnetic
  processes in dispersive media}}}\ (\bibinfo  {publisher} {Cambridge
  University Press},\ \bibinfo {year} {2005})\BibitemShut {NoStop}%
\bibitem [{\citenamefont {D'olivo}\ \emph {et~al.}(1989)\citenamefont
  {D'olivo}, \citenamefont {Nieves},\ and\ \citenamefont
  {Pal}}]{d1989electromagnetic}%
  \BibitemOpen
  \bibfield  {author} {\bibinfo {author} {\bibfnamefont {J.~C.}\ \bibnamefont
  {D'olivo}}, \bibinfo {author} {\bibfnamefont {J.~F.}\ \bibnamefont
  {Nieves}},\ and\ \bibinfo {author} {\bibfnamefont {P.~B.}\ \bibnamefont
  {Pal}},\ }\bibfield  {title} {\bibinfo {title} {Electromagnetic properties of
  neutrinos in a background of electrons},\ }\href@noop {} {\bibfield
  {journal} {\bibinfo  {journal} {Physical Review D}\ }\textbf {\bibinfo
  {volume} {40}},\ \bibinfo {pages} {3679} (\bibinfo {year}
  {1989})}\BibitemShut {NoStop}%
\bibitem [{\citenamefont {Hayes}\ and\ \citenamefont
  {Melrose}(1984)}]{hayes1984dispersion}%
  \BibitemOpen
  \bibfield  {author} {\bibinfo {author} {\bibfnamefont {L.~M.}\ \bibnamefont
  {Hayes}}\ and\ \bibinfo {author} {\bibfnamefont {D.}~\bibnamefont
  {Melrose}},\ }\bibfield  {title} {\bibinfo {title} {Dispersion in a
  relativistic quantum electron gas. i. general distribution functions},\
  }\href@noop {} {\bibfield  {journal} {\bibinfo  {journal} {Australian Journal
  of Physics}\ }\textbf {\bibinfo {volume} {37}},\ \bibinfo {pages} {615}
  (\bibinfo {year} {1984})}\BibitemShut {NoStop}%
\end{thebibliography}%

\end{document}